\documentstyle[11pt,paspconf,psfig]{article}

\def\la{\mathrel{\hbox{\rlap{\hbox{\lower4pt\hbox{$\sim$}}}\hbox{$<$}}}}
\def\ga{\mathrel{\hbox{\rlap{\hbox{\lower4pt\hbox{$\sim$}}}\hbox{$>$}}}}

\begin{document}

\title{Formation of Dark Matter Halos}

\author{J. R. Primack and J. S. Bullock}
\affil{Physics Department, University of California, Santa Cruz, CA 95964}

\author{A. A. Klypin and A. V. Kravtsov}
\affil{Astronomy Department, New Mexico State University,
Box 90001, Dept. 4500, Las Cruces, NM 88003}

\begin{abstract}
This article concerns the formation and structure of dark matter
halos, including (1) their radial density profiles, (2) their
abundance, and (3) their merger rates.  The last topic may be relevant
to the nature of the small, bright, high-redshift galaxies discovered
by the Lyman break technique.  (1) Study of a statistical sample of
galaxy-mass dark halos in high-resolution Adaptive Refinement Tree
simulations shows that they have a central density profile $\rho(r)
\propto r^{-\gamma}$ with $\gamma \approx 0.2$, in agreement with data
on dark-matter-dominated disk galaxies.  We present recent, higher
resolution results on this. (2) Another important new result is that
the Press-Schechter approximation predicts about twice as many
galaxy-mass halos at $z=0$ as are present in large dissipationless
N-body simulations; more generally, PS overpredicts the abundance of
$M \la 10^{-1} M_\ast$ halos at all redshifts.  (3) Finally, we discuss
the assembly of these halos, in particular the merger rate of
(sub-)halos at high redshift and the distribution of the starbursts
that these mergers are likely to trigger. If most of the Lyman-break
galaxies are such starbursts, this perhaps resolves the apparent
paradox that these galaxies appear to cluster like massive halos
($\sim 10^{12} M_\odot$), while their relatively low linewidths and
their spectral energy distributions suggest that they have relatively
low mass (few$\times 10^{10} M_\odot$) and young ages (few$\times
10^8$ yr). It also predicts much more star formation at high redshift
in CDM-type hierarchical models for structure formation than if only
quiescent star formation is included.
\end{abstract}

% Keywords should be included, but they are not printed in the hardcopy.


\section{Radial Profiles of Dark Matter Dominated Galaxies}

Navarro, Frenk, \& White (1995) proposed the following simple radial density
profile
\begin{equation}
\rho_{\rm NFW}(r) = {\rho_0 \over {(r/r_s)(1+r/r_s)^2}}
\end{equation}
for dark matter halos corresponding to X-ray clusters. In very
influential papers (hereafter referred to as NFW96 and NFW97) they
subsequently showed that $\rho_{\rm NFW}(r)$ is a good fit to profiles
of dark matter halos in SCDM and $\Lambda$CDM models, and in CDM
models with power law $P(k)=A k^n$ fluctuation spectra with $n=0$ to
-1.5. They characterized the halos by a concentration parameter $c
\equiv r_{200}/r_s$, where $r_{200}$ refers to the radius within which
the average overdensity is 200 times critical density.  For
$\Omega_{\rm matter}=1$ models, $r_{200}$ is approximately the same as
the virial radius. 
The mass enclosed is called $M_{200}$, and it is
useful to express this in units of the nonlinear mass $M_\ast$, which
is defined for $\Omega_{\rm matter}=1$ models by $\Delta_0(M_\ast)=
\delta_0 (1+z)$, where $\Delta_0(M)$ is the rms fluctuation of the mass
in a sphere of average mass $M$ (calculated from the linear power
spectrum) and $\delta_0=1.69$ for top-hat collapse. Finally, 
NFW96 argued that halos with lower $M_{200}/M_\ast$ 
are less concentrated because they form earlier.

The $\rho_{\rm NFW}(r)$ profile seems consistent with data on
clusters, and the $\rho \propto 1/r$ behavior at small $r$ was
consistent with earlier high-resolution N-body simulations (e.g.,
Dubinski \& Carlberg 1991).  But, 
as pointed out by Flores \& Primack (1994) and Moore (1994), and
acknowledged by NFW96, 
$\rho \propto 1/r$ for small $r$ is
inconsistent with data on dark matter dominated dwarf irregular galaxies.
Moreover, Burkert (1995) showed that the four 
galaxies considered by Moore (1994) have essentially the same
rotation curve shape, corresponding to\footnote{
The analytically implied rotation curve shape is
\begin{eqnarray}
V^2_{\rm B}(r)=2V^2_{\rm B}(r_b)
\frac{\ln[(1+r/r_b)^2(1+(r/r_b)^2)] -2{\rm tan^{-1}}[r/r_b]}{r/r_b} \nonumber,
\end{eqnarray}
where $V(r)\propto r$ for small $r$.  The peak in velocity occurs
at $r_{\rm max} \simeq 3.3 r_b$, and $V_{\rm max} = V_{\rm B}(r_{\rm max}) \simeq
 2.4V_{\rm B}(r_b)$.}
\begin{eqnarray}
\rho_{\rm B}(r) = {\rho_b \over {(1+r/r_b)[1+(r/r_b)^2]}}.
\end{eqnarray}
This makes it implausible that a complicated starburst process leading
to non-adiabatic expulsion of much of the central baryonic content of
the galaxy, such as proposed by Navarro, Eke, \& Frenk (1996), could
account for the apparent inconsistency between simulations and real
galaxies. The implausiblilty that the resolution of the discrepancy
lies in this direction was further increased when Kravtsov, Klypin,
Bullock, \& Primack (1998, hereafter KKBP98) showed that a larger set
of ten dark matter dominated dwarf irregular galaxies, and also a set
of seven dark matter dominated low surface brightness (LSB) galaxies,
all have the same rotation curve shape, again corresponding to
$\rho_{\rm B}(r)$.  This is shown in Figure 1.  In our sample we
included only those galaxies in which the dark matter component was
shown to constitute $\ga 85\%$ of the total mass inside the last
measured point of the rotation curve (in most cases with the maximum
disk assumption). 
These {\em dark matter dominated galaxies} offer a
unique opportunity for probing {\em directly} the density structure of
dark matter halos which can be then compared with predictions of
theoretical models. 
It hardly seems possible that all of these
galaxies could have had the same sort of complicated conspiracy
between dark matter and star formation, nor that the LSBs could have
lost a significant fraction of their central baryons. A more plausible
interpretation of the self-similarity of the radial mass distribution
in these dark matter dominated galaxies is that it reflects an
underlying similarity in the dark matter distribution.

McGaugh \& de Blok (1998) emphasized that the sharp $r^{1/2}$ rise in
central circular velocity predicted by the NFW $\propto r^{-1}$ density
profile is in striking disagreement with the roughly linear rise in
rotation velocity observed in LSB galaxies: ``Even treating both $c$
and $V_{200}$ as completely free parameters, no fit can be obtained.''

\begin{figure}[h]
\centerline{\psfig{file=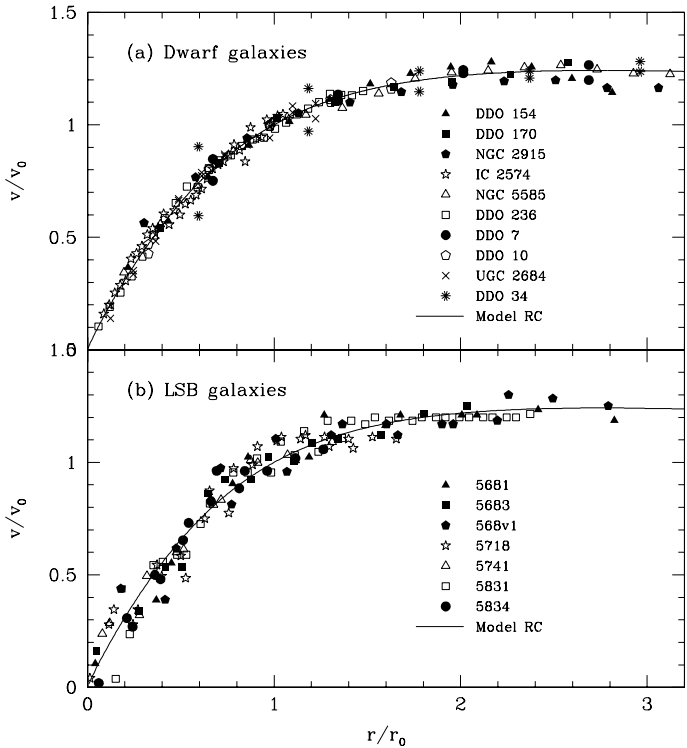,width=4.5in}}
\caption{Rotation curves of (a) ten dwarf irregular and (b) seven
low surface brightness galaxies (symbols) with measured rotation curves and
published mass models for stellar, gas, and dark matter components,
normalized to the best fit values of $r_0$ and rotational velocity $v_0$
at $r_0$ predicted by density profile $\rho_{DM}(r)$ with $(\alpha, \beta,
\gamma)=(2,3,0.2)$, which is represented by the solid line; $\rho_{\rm
B}(r)$ is nearly identical. (From KKBP98.)}
\label{fig:rcs}
\end{figure}

KKBP98 fit the observed rotation curves with a more general profile,
\begin{equation}
\rho_{DM}(r) = 
\frac{\rho_0}{(r/r_0)^{\gamma}[1+(r/r_0)^{\alpha}]^{(\beta-\gamma)/\alpha}},
{\ }\Rightarrow{\ \ }V(r)=V_t\frac{(r/r_t)^g}{[1+(r/r_t)^a]^{(g+b)/a}},
\end{equation}
in which $\rho(r \ll r_0) \propto r^{-\gamma}$, $\rho(r \gg r_0)
\propto r^{-\beta}$, and $\alpha$ characterizes the sharpness of the
change in logarithmic slope. This is equivalent to $\rho_{\rm
NFW}(r)$ for $(\alpha, \beta, \gamma)=(1,3,1)$, and to the so-called
isothermal profile with a core $r_0$ for $(\alpha, \beta,
\gamma)=(2,2,0)$. Parameters $r_t$ and $V_t$ of the corresponding
rotation curve are the effective ``turnover'' radius and velocity,
and $a$ parameterizes the sharpness of the turnover.
%\footnote{The
%limiting behaviors are $V(r \gg r_t) \propto 1/r^{b}$ and $V(r \ll
%r_t) \propto r^g$. The peak of the velocity profile occurs at the
%radius $r_{\rm max} = r_t(g/b)^{1/a}$, and $V_{max} = V(r_{max}) =
%V_t(g/b)^{g/a} [1+g/b]^{-(b+g)/a}$.} 
Because of the relatively small
range of radii probed by measured rotation curves, such measurements
cannot be used to constrain all five parameters of $\rho_{DM}$, so we
fixed the outer logarithmic slope to the value suggested by the
theoretical models $\beta=3$, $b=0.34$. The plausible value of the
parameter $\alpha=2$ was determined using galaxy rotation curves that
do show a turnover. We then found that $\gamma \approx 0.2$ fits the
observed rotation curves of our sample of 17 galaxies. The
corresponding best-fit slopes of the velocity profile are
$(a,b,g)=(1.50,0.34,0.9)$. Note that $g=1-\gamma/2$. With parameters
$\alpha$, $\beta$, and $\gamma$ (or $a,b,g$) fixed, we fitted the data
for the remaining {\em two} free parameters: $\rho_0$ and $r_0$ (or
$V_t$ and $r_t$, or in terms of Burkert's profile, $\rho_b$ and
$r_b$). The resulting structural relations show a decrease in the
characteristic density with increasing characteristic radii, or an
increase in maximum rotation velocity with increase in the radius at
which the maximum occurs.  Matching these observational relations is
a challenge for any theory that aspires to explain the observed
rotation curves.

\section{Radial Profiles From Simulations}

Does the disagreement between CDM-type simulations and the observed
rotation curves of dark matter dominated galaxies at small radii mean
that the dark matter in these galaxies is {\it not} mostly cold dark
matter?  That would be the implication if the discrepancy were real.

We have used the Adaptive Refinement Tree (ART) $N$-body code
(Kravtsov et al., 1997), which adaptively refines spatial and temporal
resolution in high density environments, to simulate the evolution of
collisionless dark matter in the three cosmological structure
formation models\footnote{The simulations followed trajectories of
$128^3$ CDM particles in a box of size of $L_{box}=7.5h^{-1}$ Mpc.
The CHDM simulation had $3\times 128^3$ particles due to the addition
of two equal-mass neutrino species.}: (a) the standard cold dark matter
(CDM) model ($\sigma_8=0.67$, $h=0.5$); (b) a low-density CDM model
with cosmological constant ($\Lambda$CDM) with parameters favored by
the high-redshift supernovae data ($\Omega_{\rm matter}=0.3$, $h=0.7$,
$\sigma_8=1$); and (c) a cold$+$hot dark matter model with two types
of neutrino (CHDM) with favorite parameters $\Omega_{\nu}=0.2$,
$h=0.5$) (Primack et al. 1995, Gawiser \& Silk 1998, but cf. Primack
\& Gross 1998).  For the dark matter halos used in KKBP98, the
spatial resolution is equal to $\approx 0.5-2h^{-1}$ kpc, and for each
of the analyzed halos we have taken into account only those regions
of the density and circular velocity profiles that correspond to
scales at least twice as large as the formal resolution.  The
reliability of the simulated density and velocity profiles was tested
by comparing results of the simulations with different resolutions and
time steps (Kravtsov et al. 1997).  A sample of $\sim 50$ halos was
extracted from each simulation, and rotation curves of halos were
fitted with the same density distribution as the data (the same set of
$\alpha$, $\beta$, and $\gamma$).  The halo rotation curves were then
renormalized to their best fit values of $r_0$ and $v_0=v(r_0)$ and
these renormalized rotation curves (i.e. $v/v_0(r/r_0)$) were averaged
over the whole sample for each model.

\begin{figure}[h]
\centerline{\psfig{file=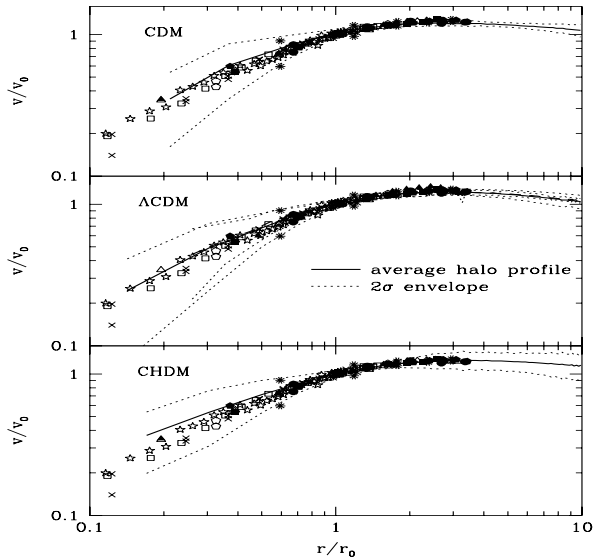,width=4.5in,height=3.15in}}
\caption{The average normalized dark matter velocity profiles for halos formed
in (a) CDM ($\sigma_8=0.67$, $h=0.5$), (b) $\Lambda$CDM ($\Omega_{\rm
matter}=0.3$, $h=0.7$, $\sigma_8=1$) and (c) CHDM ($\Omega_\nu=0.2$ in
$N_\nu=2$ neutrino species, $h=0.5$) models with corresponding profiles of
the dwarf galaxies from our sample. The dotted lines show the $2\sigma$
envelope representing scatter of individual halo profiles around the
average. Although the velocity profiles of the hierarchically formed dark
matter halos are on average consistent with the shape of observed rotation
curves, the scatter in the inner regions of the halo velocity profiles is
substantial. (From KKBP98.)}
\label{fig:ave}
\end{figure}

Figure 2 shows the average normalized dark matter velocity profiles for
halos formed in each of the three cosmological structure formation models
compared with the rotation curves of the dark matter dominated galaxies in
our sample. We find that {\em on average}, the velocity profiles of the
halos formed in the hierarchical structure formation models and observed
dark matter halos are in reasonably good agreement. Why did we not see the
significant discrepancy between numerical simulations and rotation curve
measurements indicated by previous work? One possible explanation is that
we have not used an (extrapolated) analytic model fit, but have compared
the data with the average shape of the dark matter halos directly. We also
find that dark matter dominated dwarf and LSB galaxies show structural
correlations between their characteristic density, $\rho_0$, and radius,
$r_0$, consistent with the correlations of our simulated dark matter halos:
physically smaller halos are denser. We find a similar correlation between
the maximum of the rotation curve, $v_{max}$, and the corresponding radius
$r_{max}$ (see KKBP98 for details, and also Figure 3 below).  This
increases our confidence that the agreement between the simulations and
the observed dark matter dominated galaxies is not a fluke.

In recent work, finished after KKBP98, we have run and analyzed a
$128^3$ particle CDM ART simulation ($\sigma_8=1.0$) in a box of 2.5
$h^{-1}$ Mpc, 1/3 the linear size of that used for the simulations
shown in Figure 2. Figure 3 shows the well-resolved halos fit to the
Burkert profile; as already mentioned, the $\rho_{DM}$ profile with
$\gamma \approx 0.2$ is essentially identical.  
%For each halo, the smallest radial bin is greater than twice the formal 
%resolution and contains at least ten particles.  
Again we have verified
that the structural relations, such as the correlation between the maximum
rotation velocity and the radius at which this maximum occurs, are in
reasonable agreement with our sample of dwarf and LSB galaxies --- see
Figure 4.

\begin{figure}[h]
\centerline{\psfig{file=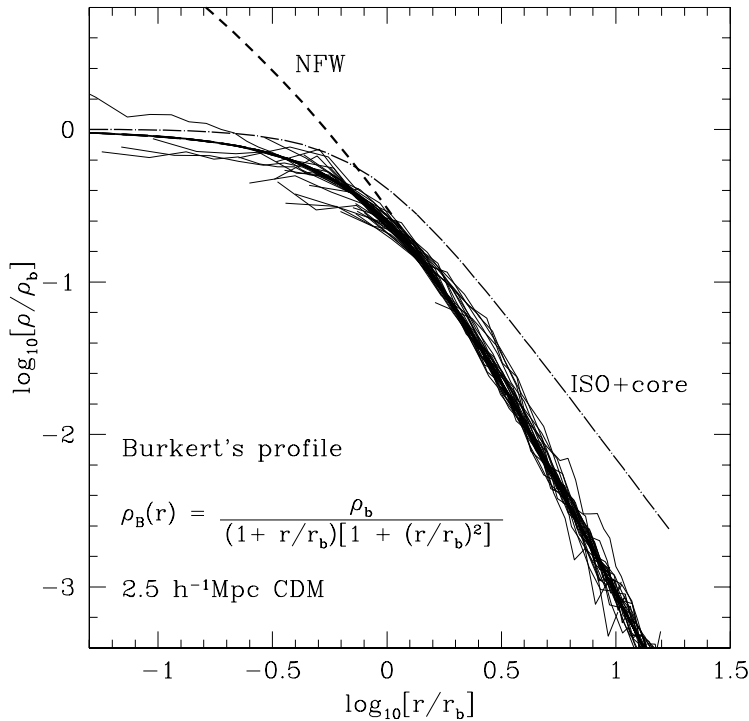,width=4.0in}}
\caption{The density profiles of halos in a CDM simulation with box
size 2.5 $h^{-1}$ Mpc, fit to Burkert's profile $\rho_B$ (solid
curve).  For each halo, the smallest radial density bin is larger
than two formal resolution elements and contains at least 10 particles.
The failure of the NFW profile (dashed curve) at small radii
is apparent, but $\rho_{\rm NFW}$ agrees well with the simulations at
larger radii where the isothermal profile with a core (dot-dash curve)
fails because it only falls as $r^{-2}$.}
\label{fig:smallbox}
\end{figure}

\begin{figure}[h]
\psfig{file=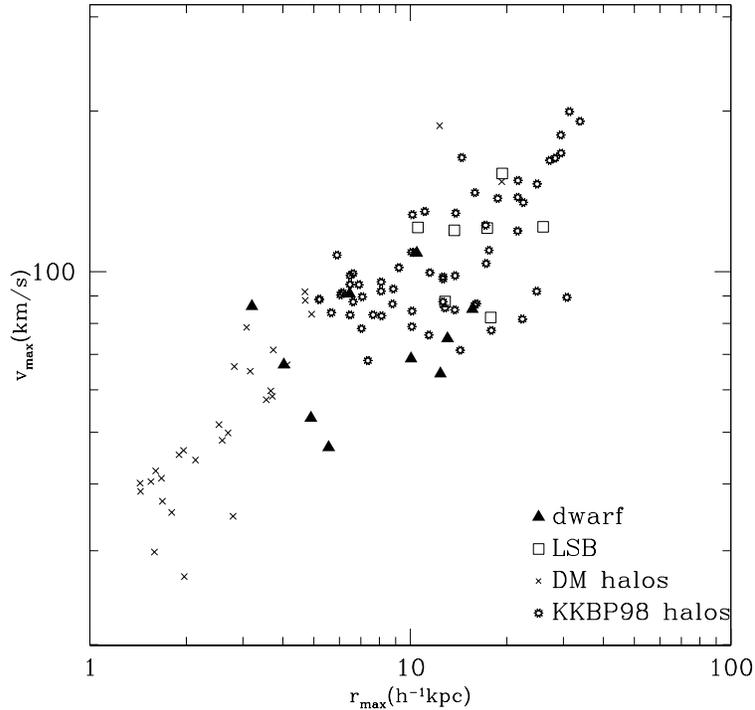,width=4.0in}
\caption{Observed versus simulated maximum rotation velocity as a function of
the radius at which it occurs.  Simulated dark matter halos are from our CDM
simulation in a box of 2.5 $h^{-1}$ Mpc (x symbols) and from the SCDM
simulation from KKBP98.  These are compared to 
$v_{\rm max}$ vs.
$r_{\rm max}$ determined from $\rho_{DM}$ fits to our sample of dwarf and
LSB galaxies.}
\label{fig:vmaxvsrmax}
\end{figure}

Moore et al. (1998) reports results from two very high resolution
simulations of clusters, which had steep central density profiles. It is
not clear that our results are in disagreement, since we consider
galaxy-mass halos and find that
% Joel, have we really found that cluster-mass halos are steeper?  I haven't
% seen that.  If anything, they are more flat -- possibly because of the
% timestep problem, and more likely because they are actually double-peaked.
%We have found that
%cluster-mass halos have steeper inner density profiles than galaxy-mass
%halos, 
%
a statistical sample of halos has a range of central
profiles.  But it is important to understand why different simulations give
different results, and in particular to understand the effects of
different resolution and different simulation techniques, of two-body
relaxation (which is suppressed in the ART approach), and of the selection
of halos to be simulated and analyzed.

It would appear to be better to compare a sample of galaxies with a
statistical sample of galaxy-mass halos, as we have done. This has many
other advantages. In work now being written up for publication, Bullock et
al. have analyzed large ART simulations at many redshifts, to study the
radial distribution of mass and angular momentum in dark matter halos at a
given epoch, and also the evolution of these properties. We find that the
dispersion of the concentration is roughly log-normal and large, and that
while it can be explained at $z=0$ by the argument presented by NFW96, this
does not work at higher redshifts. 
% added by james
In addition, halos in higher density environments tend to be more concentrated
than isolated halos, suggesting that halo selection criteria may
indeed be important for interpreting the conflicting results mentioned above.
Since the dispersion of the
concentration at $z=0$ in a given cosmology (Bullock et al. 1998) 
is larger than the difference
between different cosmologies (which comes largely from the difference
between $r_{200}$ and the true virial radius), it is incorrect to draw
conclusions about cosmology from a few galaxies (as done, for example, by
Navarro 1998, who argues that a few low-concentration galaxies favor
$\Omega_{\rm matter}$ very low).

One problem that may be resolved in the light of new data is the apparently
greater dispersion of the inner radial profiles of the simulations than of
the dark matter dominated galaxies in our sample. The work of Swaters et
al. (1997) discussed at this meeting, and also of C\^ot\'e, Freeman, \&
Carignan (1997), appears to indicate that the dispersion of the properties
of dark matter dominated galaxies may be greater than our sample suggested.
% added by james
However, as our sample in KKBP98 represented only the $\sim 50$ most massive
halos in our simulations, our scatter may be artificially small.
A major challenge to theorists remains, to explain why the dark matter
profiles of galaxies and simulations have the radial dependences
observed.  Regarding the inner profiles, Syer \& White (1998) argued
that the closer the power spectrum approximates the asymptotic CDM
slope of -3, where halos of all sizes collapse at roughly the same
epoch and therefore have the same density, the shallower the resulting
profile, while on cluster scales where smaller-mass halos collapse
earlier at higher density and are subsequently incorporated into
larger-mass halos, these will go to the center and give rise to a
steeper radial density profile. This seems consistent with our
results, as mentioned in KKBP98. We should then understand why the
results of Huss, Jain, \& Steinmetz (1998) seem inconsistent with this
argument.  Regarding why the outer radial dependence is roughly
$r^{-3}$, recent work by Henriksen \& Widrow (1998) may be relevant.

\section{Press-Schechter}

The Press-Schechter (1974, hereafter PS) formula for the number
density of dark matter halos as a function of their mass is based on
two simple assumptions -- Gaussian statistics for density
fluctuations, and spherical top-hat collapse of these fluctuations.
(See, e.g., Peebles (1993), pp. 630-635, or White (1996) for modern
treatments.)  Since both of these assumptions are known to be wrong,
or at best oversimplified, the wonder is not that the PS approximation
is not perfect, but rather that it works at all.  In fact, the PS
formula predicts the number density of virialized cluster-mass halos
in N-body simulations remarkably well.  However, several groups
recently noticed a discrepancy: $N_{PS}(>M) \approx 2 N_{\rm
simulations}(>M)$ for $M \la M_\ast/10$.  Since this has been seen in
many simulations using different methods of simulating and identifying
halos, it should be taken seriously.  For example, Gross et
al. (1998), using high-resolution particle mesh (PM) simulations and
both spherical and ellipsoidal overdensity halo finders, found that
the number density of galaxy-mass halos at the current epoch is
overestimated by PS by about a factor of 2 in many currently popular
CDM-type cosmological models, regardless of the collapse parameter
$\delta_c$ used in the PS formula.  Gross (1997), Appendix G, showed
that the discrepancy of the PS number density is about this big for
higher redshifts also, as long as $M \la M_\ast/10$.  Kauffmann et
al. (1998) found the same factor of $\sim 2$ discrepancy at the
current epoch for both the $\tau$CDM and $\Lambda$CDM models, and
Somerville, Lemson, Kolatt, \& Dekel (1998) generalized this to the
halo merging trees of the Extended PS theory.  This work is based on
AP$^3$M simulations and a friends-of-friends halo finder, as is
related work on larger mass halos, the abundance of which is
underpredicted by the PS approximation (Governato et al. 1998).
Finally, Sigad et al. (1998) have found the same phenomenon in the ART
simulations, using a version of the bound density maximum (BDM) halo
finder (Klypin et al. 1997).  A similar result was actually found
independently in an analytic calculation based on approximations
relevant to the highly nonlinear regime (Valageas \& Schaeffer 1997).

The overprediction of the number density of galaxy-mass halos at low
redshift in the PS approximation has several implications, including
an amelioration of the overprediction by semi-analytic models of
galaxy formation (based on the extended PS theory) of the luminosity
function of galaxies and the star formation rate at low redshift (see,
e.g., Somerville \& Primack 1998ab).  The underprediction by PS of the
number density of massive halos, especially at high redshifts, means
that strong conclusions about the density of the universe based on
observations of clusters compared with PS predictions should be
treated with caution (Governato et al. 1998).

\section{Lyman Break Galaxies as Merger-Triggered Starbursts}

Semi-analytic models of galaxy formation were pioneered by White \&
Frenk (1991), Kauffmann, White, \& Gurderdoni (1993), and Cole et
al. (1994); see Somerville \& Primack (1998a) for a review.  Such
models follow the evolution of the dark and luminous contents of the
universe using simple approximations to treat gas cooling, star
formation, and feedback within dark halos, and Extended PS theory to
predict the merger rate of halos in order to construct merger trees
(e.g., Somerville \& Kolatt 1998).  When halos merge, their luminous
contents are usually assumed to merge only as dynamical friction
brings smaller galaxies to the large galaxy at the center of the new
halo, and most of the star formation even at high redshift is
quiescent (e.g., Baugh et al. 1998).  However, both observational and
theoretical arguments suggest that many of the small but very bright
galaxies now being identified in very large numbers at redshifts
$z\ga3$ by the Lyman break method (Steidel et al. 1996, Dickinson
1998) are low-mass starbursts rather than large galaxies that have
been quiescently forming stars for a long time (Lowenthal et al. 1997;
Sawicki \& Yee 1998; and Somerville, Primack, \& Faber 1998, hereafter
SPF98).  SPF98 assumed that random collisions of dark matter subhalos,
as well as decay of orbits due to dynamical friction, would trigger
mergers of (proto-)galaxies that could lead to starbursts, and they
based their semi-analytic modeling of the number of such mergers on
recent dissipationless simulations of the mergers of dark halos
(Makino \& Hut 1997), and of the star formation efficiency and
timescale on hydrodynamic simulations of starbursts triggered by
mergers (Mihos \& Hernquist 1994, 1996).  The result was that most of
the star formation in CDM-type hierarchical models at redshifts $z\ga
2$ occurs in merger-triggered starbursts, and that most of the Lyman
break galaxies (LBGs) are expected to be such starbursts.  This
perhaps resolves the apparent paradox that the LBGs appear to cluster
like massive halos (Steidel et al. 1998, Giavalisco et al. 1998,
Adelberger et al. 1998; cf. Wechsler et al. 1998) while their
relatively low linewidths (Pettini et al. 1998) and their spectral
energy distributions (Sawicki \& Yee 1998) suggest that they have
relatively low mass (few$\times 10^{10} M_\odot$) and young ages
(few$\times 10^8$ yr). Including merger-triggered starbursts also
predicts much more star formation at high redshift in CDM-type
hierarchical models for structure formation than if only quiescent
star formation is included (Somerville \& Primack 1998b).

Do the merger rates actually grow so large at high redshift that this
merger-triggered starburst scenario is plausible?  With many stored
timesteps from large ART simulations, we have begun to study the
merger rate of dark matter halos as a function of redshift.  This sort
of study requires careful definitions of halos and subhalos, and of
the criteria for identifying mergers, which will be given in papers
now in preparation (Kolatt et al. 1998ab, Bullock 1999).  But to
summarize the situation briefly, the answer is yes.  The collision
rate of halos and subhalos grows in physical coordinates roughly as
$(1+z)^3$ up to a redshift that depends on the mass of the halo.  The
comoving collision rate of $\ga 10^{10} h^{-1} M_\odot$ halos peaks in
the $\Lambda$CDM simulations at $z\sim 3$, at a rate high enough to
account for the observed number density of LBGs (Kolatt et al. 1998a).
Indeed, the comoving number density of LBGs with AB magnitude brighter
than 25.5 is predicted to be almost as high at $z=4$ as at $z=3$.
We also find that the mergers occur mainly in and near the most
massive halos, so that the high bias of the bright LBGs arises
naturally.

\acknowledgments

We are grateful to our collaborators, including Avishai Dekel, Sandra
Faber, Patrik Jonsson, Tsafrir Kolatt, Michael Gross, Yair Sigad, and
Rachel Somerville, for allowing us to present preliminary reports of
our new results here.  We acknowledge support from NASA and NSF grants
at UCSC and NMSU.

\end{document}